\documentclass[a4paper,10pt,twoside,twocolumn]{article}
\usepackage{semina_tech-b}
\usepackage[latin1]{inputenc}
\usepackage[english,portuges]{babel}
\usepackage{amsmath,amssymb,amsthm}
\usepackage{graphicx}
\usepackage{times}

\begin{document}


%
\autores{R. de Oliveira}

\titulo{Renormalized Free Energy on Space-time\\
with Compact Hyperbolic Spatial Part}


 \autor{R. de Oliveira\inst{1}}

\institute{Universidade Federal do Mato Grosso (UFMT)- MT-Brasil;
rosevaldo.deoliveira@hotmail.com}

\twocolumn[ \FolhaRosto \selectlanguage{english}
\begin{abstract}
\small In this paper we found the renormalized free energy of a
interacting scalar field on a compact hyperbolic manifold
explicitly. We have shown a complete expression of the free energy
and entropy
 as a function of the curvature and the temperature. Carefully
 analyzing the free energy we have shown that there exist a minimum with
 respect to the curvature that depend on the temperature. The
 principle of minimum free energy give us an estimate of the connection between stationary curvature and
  temperature. As a result we obtain that the stationary curvature increases when the temperature increases too.
 If we start from an universe with very high curvature and temperature in the beginning, because of the principle
 of minimum free energy, the universe will reach a new situation of
 equilibrium for low temperature and low curvature. Consequently, the flat space-time is obtained for low
 temperature.
\end{abstract}

  \vspace{0.8cm}]

\section{Introduction}
The theory of quantum fields in curved space-times have been
considered by many authors$[[1],[2],[3]]$, it deals with fields in a
external gravitational field and can be considered as preliminary
step toward the complete quantum theory of gravity.

The finite temperature effective potential in quantum field theory
at finite temperature in curved space-time is the central object
which should define the behavior of the early universe.
Unfortunately it is difficult to calculate the finite temperature
potential in a general curved space-time. Therefore it is very
natural to deal with some specific spaces which are interesting from
the cosmological viewpoint.

We have considered a ultrastatic space-time because the spatial
section is a manifold with constant curvature, in such manifolds for
a fixed value of the cosmological time, it describe locally
spatially homogeneous isotropic universes [4]. The compact
hyperbolic manifolds also have the constant curvature spatial
section, but the spectrum of the relevant operator is, in general,
not explicitly known \footnote{ When the eigenvalues are known, the
zeta function can be computed explicitly [see for example ref.
[5]]}. However, there exist a mathematical tool, the Selberg trace
formula [4], [6], [7] and [8], which permits to evaluate some
interesting physically global quantities, like the vacuum energy or
the one-loop finite and zero temperature effective potential.

Throughout the paper we will use the zeta-function technique for
regularizing the path integral [4],[5],[21] and [22].

In this work we will consider a model of a self-interacting scalar
field on a curved space-time with compact hyperbolic spatial part
[13].

The finite temperature effects of this model are studied by
employing a complex integral representation for the finite
temperature potential (or free energy) [13]-[19],[22].

Taking into account the principle of minimum free energy (or maximum
of entropy) and analyzing the behavior of the renormalized free
energy, we can deduce the connection between stationary curvature
with the
 temperature.

\section{Finite and zero temperature effective Potential}

Before we restrict our attention to the hyperbolic manifolds ${\cal
M}=S^{1} \times \Re^{1}  \times H^{2}/\Gamma $, let us define the
zero and finite temperature effective potential. The  Euclidean
generating functional for Greens functions in a scalar field theory
is given by, see ref. [1]

\begin{eqnarray}
Z_{E}[J]=e^{-\frac{W_{E}[J]}{\hbar}}=N_{E}\int {\cal D}\varphi
e^{-\frac{S_{E}[\varphi,J]}{\hbar}},
\end{eqnarray}

\noindent where $W_{E}[J]$ generates connected Greens functions. The
Euclidean action is

\begin{eqnarray} 
S_{E}[\varphi,J]&=&\int d^{4} x
\left[\frac{1}{2}\partial_{\mu}\varphi\partial_{\mu}\varphi+\frac{1}{2}
m^{2} \varphi^{2} + V(\varphi) \nonumber \right.\\
-\left.J \varphi\right].&&
\end{eqnarray}

Let us define the following relation

\begin{eqnarray} 
\frac{\delta W_{E}[J]}{\delta J(x)}=-\phi_{c}(x),
\end{eqnarray}

\noindent $\phi_{c}(x)$ is a classical variable.

The effective action is defined by

\begin{eqnarray} 
\Gamma_{E}[\phi_{c}]=W_{E}[J]+\int dx \phi_{c}(x) J(x).
\end{eqnarray}

In the one-loop approximation we find that the Euclidean generate
functional is given by

\begin{eqnarray} 
Z_{E}[J]= e^{\frac{-S_{E}[\phi_{c},J]}{\hbar}}\left[\det
S^{E}_{2}\mu^{2}\right]^{-\frac{1}{2}},
\end{eqnarray}

\noindent where

\begin{eqnarray} 
S_{2}^{E}=\left\{-\partial_{\mu}\partial_{\mu}+m^{2} +
V''(\phi_{c})\right\}\delta(x_{1}-x_{2}).
\end{eqnarray}

\noindent the  constant $\mu$ is introduced to become the logarithm
dimensionless. The above expression can be proved to have only
diagrams with one independent integral over the four momenta.

\subsection{ The zero temperature effective potential }

We expect to be able to write $\Gamma_{E}[\phi_{c}]$ in the form

\begin{eqnarray} 
\Gamma_{E}[\phi_{c}]&=&\int
dx\left(-V_{eff}(\phi_{c}(x))+\right.\nonumber\\&&\left.\frac{1}{2}A(\phi_{c}(x))\partial_{\mu}\phi_{c}(x)\partial_{\mu}\phi_{c}(x)+\ldots\right).
\end{eqnarray}

If we set the source equal to zero, then the classical field takes a
constant value $\phi_{c}(x)=\phi_{c}$\footnote{You can see looking
at in the equation $\frac{\delta \Gamma[\phi_{c}]}{\delta
\phi_{c}}=0$}.
In this limit the effective action is given to be
\begin{eqnarray} 
\Gamma_{E}[\phi_{c}]=V_{E} V_{eff}(\phi_{c}),
\end{eqnarray}
\noindent whre $V_{E}$ is the Euclidian measure. The effective
action is written as\footnote{ we need to remember that$$\left[\ln
(Z_{E}[J])-\ln \left(e^{-\frac{S_{E}[\phi_{c},J]}{\hbar}}\left[\det
A\mu^{2}\right]^{-\frac{1}{2}}\right)\right]_{J=0,\phi_{c}=cte}=0
$$}
\begin{eqnarray} 
\Gamma_{E}[\phi]&=&S_{E}[\phi]+\hbar \bar{\Gamma}^{1}_{E}[\phi_{c}]\\
V_{eff}&=&V_{cl}+\frac{\hbar}{V_{E}}\bar{\Gamma}^{1}_{E}[\phi_{c}]
\nonumber
\end{eqnarray}
Using the Eq.[11] we can write the effective potential as
\begin{eqnarray} 
V_{eff}=V_{cl}+\frac{\hbar}{2 V_{E}}\ln\left[\det
S^{E}_{2}\mu^{2}\right].
\end{eqnarray}

\subsection{The finite temperature effective potential }

The theory at finite temperature, $T=\frac{1}{\beta}$, can be
obtained by compactifying the imaginary time $\tau$  and assuming
periodic conditions on the fields,
$\phi(\tau,x)=\phi(\tau+\beta,x)$. The partition function of a
scalar fields in thermal equilibrium at finite temperature $T$,(see
ref.[2]) is defined as
\begin{eqnarray} 
Z_{\beta}=e^{-\frac{W_{\beta}[J]}{\hbar}}=N_{E}\int {\cal D}\varphi
e^{\frac{- S_{E}[\varphi,J]}{\hbar}},
\end{eqnarray}
\noindent where the field satisfy a periodic boundary condition
$\varphi(\tau+\beta)=\varphi(\tau)$. We define the following
relation
\begin{eqnarray}
\phi_{c}(x)=-\frac{\delta W_{\beta}[J]}{\delta
J(x)}=\frac{\hbar}{Z_{\beta}[J]}\frac{\delta Z_{\beta}[J]}{\delta
J(x)}=<\hat{\phi}(x)>_{J}
\end{eqnarray}
The finite temperature effective action is defined as
\begin{eqnarray}
\Gamma_{\beta}[\phi_{c}]=W_{\beta}[J]+\int_{0}^{\beta}d\tau\int dx
J(x)\phi_{c}(x)
\end{eqnarray}
By using of the result of the last section, to one loop
approximation we can write the partition function as
\begin{eqnarray} 
Z_{\beta}= e^{\frac{-S_{E}[\phi_{c},J]}{\hbar}}\left[\det
S^{E}_{2,\beta}\mu^{2}\right]^{-\frac{1}{2}}
\end{eqnarray}
The free energy is given to be
\begin{eqnarray} 
F=-\frac{1}{\beta}\ln
Z_{\beta}=\frac{S_{E}}{\hbar\beta}+\frac{1}{2\beta}\left[\det
S^{E}_{2,\beta}\mu^{2}\right]
\end{eqnarray}
When $\phi_{c}(x)=cte$ the free energy per unit volume is
\begin{eqnarray} 
\frac{\hbar F}{\Omega}=V_{c}+\frac{\hbar}{2 \Omega \beta}\ln
\left[\det S^{E}_{2,\beta}\mu^{2}\right]
\end{eqnarray}
\noindent where $V_{E}=\beta\Omega$. Thus we define the finite
effective potential as
\begin{eqnarray} 
V_{eff}^{\beta}&=&\frac{\hbar F}{\Omega}=V_{c}+\frac{\hbar}{2 \Omega \beta}\ln \left[\det S^{E}_{2,\beta}\mu^{2}\right]\\
V_{eff}^{\beta}&=&V_{c}+\frac{\hbar F_{0}}{\Omega}+\frac{\hbar F_{\beta}}{\Omega}\nonumber.\\
V_{c}&=&\frac{\lambda
\phi^{4}}{4!}+\frac{m^{2}\phi^{2}}{2}+\frac{\xi R \phi^{2}}{2}.
\end{eqnarray}

\section{One-Loop Effective Potential on ${\cal M}=S^{1}\times R^{1}\times H^{2}/\Gamma$}
With the purpose of find the one-loop effective potential, we will
consider the representation of the zeta-function given by
\begin{eqnarray} 
&&\frac{\zeta(s|A)}{\Omega_{1}}=\frac{\beta\Gamma(s-1)}{4\pi\Gamma(s)}\zeta(s-1|L_{2})\\&&
+\frac{1}{4\pi\Gamma(s)\pi i}\int_{\Re \!
z=C}\zeta_{R}(z)\Gamma(\frac{z-1}{2}+s-\frac{1}{2})\nonumber\\
&&\times\Gamma(\frac{z}{2})\left(\frac{\beta}{2}\right)^{-z+1}\zeta(\frac{z-1}{2}+s-\frac{1}{2}|L_{2})dz.\nonumber
\end{eqnarray}
The one-loop effective potential is given by Eq(10), i.e
\begin{eqnarray} 
V_{eff}=V_{c}+V^{(1)}.
\end{eqnarray}
We know that the determinant of the operator is relatad to
zeta-function as
\begin{eqnarray} 
&&V^{(1)}=\frac{1}{2\Omega({\cal M})}\ln\det(A\mu^{2})\nonumber\\
&&=\frac{1}{2\Omega({\cal M})}\lim_{s\to
0}\left[-\frac{d\zeta(s|A)}{ds}+\zeta(s|A)\ln\mu^{2}\right].
\end{eqnarray}
Now we can write the quantum correction by using the above results
as
\begin{eqnarray} 
&&V^{(1)}=V_{0}^{(1)}+V^{(1)}_{\beta}\\
&&=\frac{1}{2\Omega({\cal M})} \left\{ +\frac{\beta}{4\pi\Omega_{1}}\left[ R_{H}^{-2} {\cal F}(0;\delta)\right.\right.\nonumber\\
&&\left.\left. + \lim_{s\to 0} \frac{d}{ds}\left( R_{H}^{2(s-1)}{\cal F}(s;\delta)\right) \right]\right. \nonumber\\
&&- \left.\frac{1}{4\pi^{2} i\Omega_{1}} \int_{\Re z=C}
\zeta_{R}(z)\Gamma(\frac{z}{2}) \Gamma\left(\frac{z-2}{2}\right)
 R_{H}^{z-2}{\cal F}(\frac{z}{2};\delta)\right.\nonumber\\
 &&\left.\left(\frac{\beta}{2}\right)^{-z+1} dz-\frac{\beta\Omega_{1}}{4\pi} R_{H}^{-2}{\cal F}(0;\delta)\ln\mu^{2}
\right\}\nonumber
\end{eqnarray}
The volume is $\Omega({\cal M})=\beta\Omega_{1} R_{H}^{2} V({\cal
F})$. In order to derive the zero temperature contribution we need
to calculate the derivative of the Bytsenko function
\begin{eqnarray} 
&&\frac{d{\cal F}(s;\delta)}{ds}=\frac{V({\cal F})}{4\pi}\left[\frac{\delta^{2(2-s)}(-2)\ln\delta}{(s-2)}+\frac{\delta^{2(2-s)}(-1)}{(s-2)^{2}}\right]\nonumber\\
&&-\frac{V({\cal F})}{2\pi}\displaystyle\frac{dL_{1}(\Delta^{1-s})}{ds}+\left\{H(s)\left[\frac{-\psi(s-1)}{\Gamma(s-1)\Gamma(2-s)}\right.\right.\\
&&\left.\left.+\frac{\psi(2-s)}{\Gamma(s-1)\Gamma(2-s)}\right]+\frac{H^{\prime}(s)}{\Gamma(s)\Gamma(2-s)}\right\} \nonumber\\
&&+T_{0}\displaystyle\frac{dL_{3}(\Delta^{1-s})}{ds}\end{eqnarray}
\noindent Where we have defined
\begin{eqnarray} 
L_{1}(\Delta^{1-s}) &=& \int_{0}^{\infty}  r (1-\tanh \pi r)(r^{2} + \delta^{2})^{1-s} dr\nonumber\\
&=& 2\int_{0}^{\infty} \frac{r(r^{2}+ \delta^{2})^{1-s}}{1+ e^{2\pi
r}} dr,
\end{eqnarray}
\begin{eqnarray} 
L_{3}(\Delta^{1-s}) = \int_{-\infty}^{+\infty} \frac{e^-\frac{2\pi k
r}{n}}{1+ e^{-2\pi r}}(r^{2} + \delta^{2})^{1-s} dr,
\end{eqnarray}
\begin{eqnarray} 
H(s)=\int_{0}^{\infty} dy
\frac{Z^{\prime}(y+\delta+1/2)}{Z(y+\delta+1/2)}(y^{2}+2y\delta)^{1-s},
\end{eqnarray}
\noindent where $\Delta=\delta^{2}+r^{2}$ and
$T_{0}=\sum_{\{Q\}}\sum_{k=1}^{m-1}\frac{\xi^{k}(Q)}{2m\sin(\frac{k\pi}{m})}$.
The zero temperature contribution to effective potential is written
as
\begin{eqnarray} 
&&V_{0}^{(1)}=-\frac{1}{128\pi^{2}}\left\{\frac{\Lambda^{2}}{2}\left[\frac{3}{2}+\ln\left(\frac{2}{\Lambda\mu^{2}}\right)\right]\right.\nonumber\\
&&-2F_{3}(R,\phi)+\frac{4\pi R^{2} H(0)}{V({\cal F})}+\left.R\left[\frac{\Lambda}{12}+\frac{7R}{480}\right]\right.\nonumber\\
&&\left. \left[1+\ln\left(\frac{2}{\Lambda\mu^{2}}\right)\right]\right\}-\frac{1}{64\pi V({\cal F})}\times\nonumber\\&&\times\sum_{\{Q\}}\sum_{k=1}^{m-1}\left\{-f(k)R\left[\Lambda+Rh(k)\right] \left[1+\ln\left(\frac{2}{\Lambda\mu^{2}}\right)\right]\right.\nonumber\\
&&\left.+g(k)G_{3}(R,\phi)\right\}
\end{eqnarray}
\noindent here we have used
\begin{eqnarray} 
f(k)&=&\frac{\xi^{k}(Q)} {2m\sin^{2}\left(\frac{k\pi}{m}\right)}\\
g(k)&=&\frac{\xi^{k}(Q)} {m\sin\left(\frac{k\pi}{m}\right)}\nonumber\\
F_{3}(R,\phi)&=&\int_{0}^{\infty} dr\frac{2r} {1+e^{2\pi r}} R(Rr^{2}+\Lambda)\times\nonumber\\&&\times\ln\left(1+\frac{Rr^{2}}{\Lambda}\right)\nonumber\\
G_{3}(R,\phi)&=&\int_{-\infty}^{+\infty} dr \frac{e^{-\frac{2\pi kr}{m}}}{1+e^{-2\pi r}} R(Rr^{2}+\Lambda)\nonumber\\&&\times\ln\left(1+\frac{Rr^{2}}{\Lambda}\right)\nonumber\\
\Lambda&=&R\delta^{2}\nonumber\\
h(k)&=&-\frac{1}{4}+\frac{1}{2}\csc^{2}(\frac{k\pi}{m})\nonumber.\\
&&R\to-R\nonumber.
\end{eqnarray}
Now we are going to calculate the finite temperature contribution to
quantum corretion of the effective potential
\begin{eqnarray} 
&&V_{\beta}^{(1)}=-\frac{R_{H}^{-4}}{16\pi^{2}V({\cal F})i}\int_{Re \!\! z=C}\zeta_{R}(z)\Gamma(z/2)\nonumber\\&&\times\Gamma\left(\frac{z}{2}-1\right)\left(\frac{\beta}{2}\right)^{-z} R_{H}^{z}\left\{ \frac{V({\cal F})\delta^{2(2-\frac{z}{2})}}{4\pi(\frac{z}{2}-2)} \right.\nonumber\\
&&- \frac{V({\cal F})}{2\pi}\int_{0}^{\infty} r(1-\tanh \pi r)(r^{2}+\delta^{2})^{1-\frac{z}{2}}dr \nonumber\\
&&+
\frac{(2\delta)^{\frac{3}{2}-\frac{z}{2}}}{\sqrt{\pi}\Gamma(\frac{z}{2}-1)}\sum_{\{P\}_{P}}
\sum_{k=1}^{\infty}{\cal
A}(k,P)[kl(P)]^{\frac{z}{2}-\frac{3}{2}}\nonumber\\
&&\times K_{\frac{z}{2}-\frac{3}{2}}[\delta k l(P)] +
\sum_{\{Q\}}\sum_{k=1}^{n-1} B(Q,k)\\&&\times
\int_{-\infty}^{+\infty}\left. \frac{ e^{-\frac{2\pi r
k}{n}}}{1+e^{-2\pi r}}
(r^{2}+\delta^{2})^{1-\frac{z}{2}}dr\right\}dz \nonumber
\end{eqnarray}
After the calculation of the above integrals we find that
\begin{eqnarray} 
V_{\beta}^{(1)}=V_{\beta}^{I}+V_{\beta}^{h}+V_{\beta}^{\epsilon}
\end{eqnarray}
\noindent where we have
\begin{eqnarray} 
&&V_{\beta}^{I}=-\frac{\pi^{2}\beta^{-4}}{90}+\frac{\Lambda\beta^{-2}}{48}+\frac{R\beta^{-2}}{576}-\frac{\Lambda^{\frac{3}{2}}\beta^{-1}}{24\sqrt{2}\pi}\\
&&-\frac{F_{1}(R,\phi)\beta^{-1}}{8\pi\sqrt{2}}+\frac{1}{128\pi^{2}}\left\{\frac{\Lambda^{2}}{2}\left[\frac{3}{2}+\ln\left(\frac{32\pi^{2}}{\beta^{2}\Lambda}\right)\right]\right.\nonumber\\
&&\left.+R\left[\frac{\Lambda}{12}+\frac{7R}{480}\right]\left[1+\ln\left(\frac{32\pi^{2}}{\beta^{2}\Lambda}\right)\right]\right.\nonumber\\
&&\left.-\Lambda^{2}\gamma-2R\gamma\left[\frac{\Lambda}{12}+\frac{7R}{480}\right]-2F_{3}(R,\phi)\right\}     \nonumber\\
&&+\frac{S\Lambda^{n+1}R^{n+1}\beta^{2n-2}}{16\pi^{2}}+\frac{S_{1}F_{n}(R,\phi)\beta^{2n-2}}{16\pi^{2}2^{3n-1}}\nonumber
\end{eqnarray}
\begin{eqnarray} 
V_{\beta}^{h}=\frac{R^{2}T_{1}I_{1}(R,\phi)}{32\pi V({\cal
F})}-\frac{R^{\frac{3}{2}}T_{1}
I_{2}(R,\phi)\beta^{-1}}{4\sqrt{2}\pi V({\cal F})}
\end{eqnarray}
\begin{eqnarray} 
&&V_{\beta}^{\epsilon}=-\frac{1}{32\pi V({\cal F})}\sum_{\{Q\}}\sum_{k=1}^{m-1}\left\{\frac{4f(k)R\pi^{2}\beta^{-2}}{3}\right.\\
&&\left.-2^{\frac{3}{2}}\pi g(k) G_{1}(R,\phi)\beta^{-1}+\frac{f(k)}{2}[R\Lambda+R^{2}h_{k}]\right.\\
&&\left.\times[1-2\gamma+\ln\left(\frac{32\pi^{2}}{\Lambda \beta^{2}}\right)-\frac{g(k)}{2}G_{3}(R,\phi)\right\}   \nonumber\\
&&+\frac{T_{0}S_{1}G_{n}(R,\phi)\beta^{2n-2}}{32\pi
2^{3n-3}}\nonumber
\end{eqnarray}
\noindent where we have defined the following constants
\begin{eqnarray} 
F_{1}(R,\phi)&=&\int_{0}^{\infty} dr\frac{2r} {1+e^{2\pi r}} R(Rr^{2}+\Lambda)^{\frac{1}{2}}\\
G_{1}(R,\phi)&=&\int_{-\infty}^{+\infty} dr \frac{e^{-\frac{2\pi kr}{m}}}{1+e^{-2\pi r}} R(Rr^{2}+\Lambda)^{\frac{1}{2}}\nonumber\\
I_{1}(R,\phi)&=&\int_{0}^{\infty}dy e^{-(y+\delta)kl(P)}(y^{2}+2y\delta)\nonumber\\
I_{2}(R,\phi)&=&\int_{0}^{\infty}dy e^{-(y+\delta)kl(P)}(y^{2}+2y\delta)^{\frac{1}{2}}\nonumber\\
F_{n}(R,\phi)&=&\int_{0}^{\infty} dr\frac{2r} {1+e^{2\pi r}} R(Rr^{2}+\Lambda)^{n}\nonumber\\
G_{n}(R,\phi)&=&\int_{-\infty}^{+\infty} dr \frac{e^{-\frac{2\pi
kr}{m}}}{1+e^{-2\pi r}} R^{2n}(Rr^{2}+\Lambda)^{1-n}\nonumber\\
T_{1}&=&\sum_{\{P\}_{\Gamma}}\sum_{k=1}^{\infty}{\cal
A}(k,P)\nonumber\\
T_{0}&=&\sum_{\{Q\}}\sum_{k=1}^{m-1}{\cal
B}(k,Q)\nonumber\\
S&=&\sum_{n=2}^{\infty}\frac{(-1)^{n}\pi^{\frac{3}{2}-2n}\Gamma(n-\frac{1}{2})\zeta(2n-1)}{(n+1)!2^{3n-1}}\nonumber\\
S_{1}&=&\sum_{n=2}^{\infty}\frac{2(-1)^{n}\pi^{\frac{3}{2}-2n}\Gamma(n-\frac{1}{2})\zeta(2n-1)}{(n)!2^{3n-1}}\nonumber
\end{eqnarray}


\section{Renormalization of zero temperature effective potential}

For  simplicity we will consider the massless case $m=0$. The
renormalized zero temperature effective potential is defined by
\begin{eqnarray} 
V_{0}^{R}=V_{c}+V^{(1)}_{0}+V_{a}+\delta V
\end{eqnarray}
\noindent where
\begin{eqnarray} 
\delta V&=&\frac{\delta\lambda\phi^{4}}{4!}+\frac{\delta\xi R \phi^{2}}{2}+\frac{\delta a R^{2}}{2}\\
V_{a}&=&\frac{a R^{2}}{2}\nonumber
\end{eqnarray}
The counterterms are fixed by the following renormalization
conditions \cite{andrei-Rep}
\begin{eqnarray} 
\delta\xi&=&-\left.\frac{\partial^{3}V_{0}^{(1)}}{\partial R\partial\phi^{2}}\right|_{R=0,\phi=M}\\
\delta a&=&-\left.\frac{\partial^{2}V_{0}^{(1)}}{\partial
R^{2}}\right|_{R=0,\phi=M}\\
\delta\lambda&=&-\left.\frac{\partial^{4}V_{0}^{(1)}}{\partial\phi^{4}}\right|_{R=0,\phi=M}
\end{eqnarray}
The counterterms were computed explicitly, they are
\begin{eqnarray}
\delta\lambda=-\frac{{\lambda}^{2}}{32\pi^{2}}\, \left[ 8+3\,\ln
\left( {\frac {\lambda\,{\phi}^{2}{\mu}^{2}}{2}}
 \right)  \right]
\end{eqnarray}
\begin{eqnarray}
&&\delta a=-{\frac {1}{192}}\, \left\{
2\,\lambda\,V_{{F}}+6\,\lambda\,\xi\,\ln
 \left( 2 \right) V_{{F}}\right.\nonumber\\
&&\left.+6\,\lambda\,\xi\,\ln  \left( {\frac {1}{
\lambda\,{M}^{2}{\mu}^{2}}} \right) V_{{F}}-12\,\lambda\,\xi\,V_{{F
}}+6\,\lambda\,f_{{k}}\ln  \left( 2 \right) \pi \right.\nonumber\\
&&\left.+6\,\lambda\,f_{{k}} \ln  \left( {\frac
{1}{\lambda\,{M}^{2}{\mu}^{2}}} \right)
 \pi\right.\nonumber\\
&& - \left.\lambda\,\ln  \left( 2 \right) V_{{F}}-\lambda\,\ln
\left( {\frac {1} {\lambda\,{M}^{2}{\mu}^{2}}} \right)
V_{{F}}\right.\nonumber\\
&&\left.-12\,\lambda\,f_{{k}} \pi  \right\} {\pi
}^{-2}{V_{{F}}}^{-1}
\end{eqnarray}
\begin{eqnarray}
\delta\xi&=&-\frac{\lambda}{32
\pi^{2}}\left[2+\ln\left(\frac{\lambda
M^{2}\mu^{2}}{2}\right)\right]\nonumber\\&&\times\left[\frac{1}{6}-\xi-\frac{f_{k}\pi}{V_{F}}\right]
\end{eqnarray}
\noindent where we have used the following notation
\begin{eqnarray} 
F_{3}&\equiv&\int_{0}^{\infty} dr\frac{2r} {1+e^{2\pi r}}\\
G_{3}&\equiv&\int_{-\infty}^{+\infty} dr \frac{e^{-\frac{2\pi kr}{m}}}{1+e^{-2\pi r}}\nonumber\\
f_{k}&\equiv&\frac{\xi^{k}(Q)} {2m\sin^{2}\left(\frac{k\pi}{m}\right)}\nonumber\\
g_{k}&\equiv&\frac{\xi^{k}(Q)}
{m\sin\left(\frac{k\pi}{m}\right)}\nonumber\\
J&=&\csc^{2}(\frac{k\pi}{m})\nonumber
\end{eqnarray}
Thus, the renormalized zero effective potential is a very
complicated expression
\begin{eqnarray} 
V_{0}^{R}&=&V_{C}+\frac{a R^{2}}{2}+V_{0}^{(1)}+\frac{\delta\lambda \phi^{4}}{4!}+\frac{\delta\xi R \phi^{2}}{2}+\frac{\delta a R^{2}}{2}\nonumber\\
&=&a R^{2}+b R \phi^{2}+c \phi^{4}+d+(e R^{2}+f R \phi^{2}+g
\phi^{4})\nonumber\\&&\times\ln\left[\frac{\lambda M^{4}}{\lambda
\phi^{2}-2\xi R+\frac{R}{4}}\right]
\end{eqnarray}
\noindent where $a,b,c,d,e,f$ and $g$ are constants.
\subsection{Linear curvature approximation}
Taking into account only linear terms in curvature, we find the
following result
\begin{eqnarray}
V_{0}^{R}&\approx&\frac{\lambda \phi^{4}}{4!}-\frac{1}{2}\xi R
\phi^{2}+\frac{\lambda^{2}\phi^{4}}{256 \pi^{2}}\left[\ln(\frac{\phi^{2}}{M^{2}})-\frac{25}{6}\right]\nonumber\\
&&-\frac{\lambda R \phi^{2}}{64
\pi^{2}V_{F}}\left[\ln(\frac{\phi^{2}}{M^{2}})-3\right](\xi-\frac{1}{6}+f_{k})
\end{eqnarray}
\noindent  the expression above agree with the result of the
reference [13]. It is very ease to check that when the curvature is
zero $(R=0)$ we obtain the well know result of S. Coleman and E.
Weinberg [24]
\begin{eqnarray}
V_{0}^{R}\approx\frac{\lambda
\phi^{4}}{4!}+\frac{\lambda^{2}\phi^{4}}{256
\pi^{2}}\left[\ln(\frac{\phi^{2}}{M^{2}})-\frac{25}{6}\right]
\end{eqnarray}
\section{Renormalized Finite Temperature Effective Potential}
All the ambiguities  of the effective potential are in the zero
temperature effective potential $(V_{0})$. Therefore we must define
the finite temperature effective potential in a way that the finite
temperature part is not affected by the renormalization processes,
namely, when the temperature is equal zero ($T=0)$ the finite
effective potential goes to the zero temperature effective
potential. Following the above considerations the finite temperature
effective potential is defined by
\begin{eqnarray}
V_{eff}^{R}=V_{0}^{R}+V_{\beta}^{(1)}
\end{eqnarray}
In writing the above expression becomes
\begin{eqnarray}
&&V_{eff}^{R}=a R^{2}+b R \phi^{2}+c \phi^{4}+d+(e R^{2}+f R
\phi^{2}\nonumber\\
&&+g \phi^{4})\ln\left[\frac{\lambda M^{4}}{\lambda
\phi^{2}-2\xi R+\frac{R}{4}}\right]-\frac{\pi^{2}\beta^{-4}}{90}+\frac{\Lambda\beta^{-2}}{48}\nonumber\\
&&+\frac{R\beta^{-2}}{576}-\frac{\Lambda^{\frac{3}{2}}\beta^{-1}}{24\sqrt{2}\pi}-\frac{F_{1}(R,\phi)\beta^{-1}}{8\pi\sqrt{2}}\nonumber\\
&&+\frac{1}{128\pi^{2}}\left\{\frac{\Lambda^{2}}{2}\left[\frac{3}{2}+\ln\left(\frac{32\pi^{2}}{\beta^{2}\Lambda}\right)\right]\right.\nonumber\\
&&\left.+R\left[\frac{\Lambda}{12}+\frac{7R}{480}\right]\left[1+\ln\left(\frac{32\pi^{2}}{\beta^{2}\Lambda}\right)\right]\right.\nonumber\\
&&\left.-\Lambda^{2}\gamma-2R\gamma\left[\frac{\Lambda}{12}+\frac{7R}{480}\right]-2F_{3}(R,\phi)\right\}     \nonumber\\
&&+\frac{S\Lambda^{n+1}R^{n+1}\beta^{2n-2}}{16\pi^{2}}+\frac{S_{1}F_{n}(R,\phi)\beta^{2n-2}}{16\pi^{2}2^{3n-1}}\nonumber\\
&&+\frac{R^{2}T_{1}I_{1}(R,\phi)}{32\pi V({\cal
F})}-\frac{R^{\frac{3}{2}}T_{1}
I_{2}(R,\phi)\beta^{-1}}{4\sqrt{2}\pi V({\cal F})}\nonumber\\
&&-\frac{1}{32\pi V({\cal F})}\sum_{\{Q\}}\sum_{k=1}^{m-1}\left\{\frac{4f(k)R\pi^{2}\beta^{-2}}{3}\nonumber\right.\\&&\left.-2^{\frac{3}{2}}\pi g(k) G_{1}(R,\phi)\beta^{-1}+\frac{f(k)}{2}[R\Lambda+R^{2}h(k)]\right.\nonumber\\
&&\left.\times[1-2\gamma+\ln\left(\frac{32\pi^{2}}{\Lambda \beta^{2}}\right)-\frac{g(k)}{2}G_{3}(R,\phi)\right\}   \nonumber\\
&&+\frac{T_{0}S_{1}G_{n}(R,\phi)\beta^{2n-2}}{32\pi 2^{3n-3}}
\end{eqnarray}

\subsection{High Temperature Limit}

The limit of high temperature $(\beta\to 0)$, and when the curvature
is equal zero $(R=0)$ we find the result of the literature
\begin{eqnarray}
&&V_{eff}^{R}=c \phi^{4}+d+g\phi^{4}\ln\left[\frac{\lambda
M^{4}}{\lambda
\phi^{2}}\right]-\frac{\pi^{2}\beta^{-4}}{90}+\frac{\Lambda\beta^{-2}}{48}\nonumber\\
&&-\frac{\Lambda^{\frac{3}{2}}\beta^{-1}}{24\sqrt{2}\pi}+\frac{1}{128\pi^{2}}\left\{\frac{\Lambda^{2}}{2}
\left[\frac{3}{2}+\ln\left(\frac{32\pi^{2}}{\beta^{2}\Lambda}\right)\right]-\Lambda^{2}\gamma\right\}\nonumber\\
\end{eqnarray}

\noindent less than a constant that does not depend of the
temperature, the above limit result agree with  the literature again
[23].

\section{The behavior of the Free Energy}
The free energy $F$ is defined  in term of the effective potential
as
\begin{eqnarray}
F=\frac{\Omega}{\hbar}V_{eff}^{R}.
\end{eqnarray}
We can define the order parameter $(<0|\hat{\phi}(x)|0>=\phi)$ by
the following extreme condition
\begin{eqnarray}
\left.\frac{\partial F}{\partial\phi}\right|_{\phi=\phi_{0}}=0
\end{eqnarray}
By mean of this equation we can eliminate the order parameter by
chosen the solutions to positive temperature and curvature. From
above expression we find two different solutions to the order
parameter, they are solutions very complicated, found using the
Mathematica-5 program. When we substitute these two solutions in the
free energy, we observe that the behavior of the free energies are
the same to them. Furthermore, none of the roots intercept
themselves. Therefore, we chose the solution of less energy.

The next step is to study the behavior of the free energy of our
model. For a constant value of the temperature, the behavior of the
free energy for low curvature values is shown in the graph of {\bf
Figure-1}.
\begin{figure}[h]
\begin{center}
\includegraphics[width=0.5\textwidth]{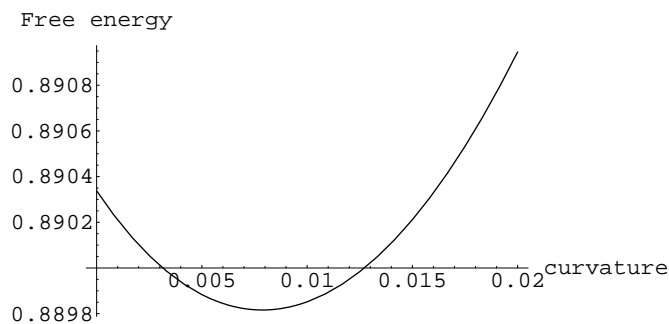}
\end{center}
\caption{The behavior of the entropy with respect to the curvature
for a constant value of the temperature}
\end{figure}
The behavior of the free energy with respect to the temperature for
a constant value of the curvature can be seen in the graph of the
{\bf Figure-2}.
\begin{figure}[h]
\includegraphics[width=0.5\textwidth]{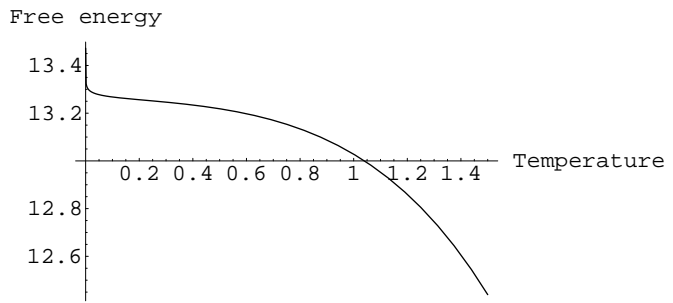}
\caption{The behavior of the free energy with respect to the
temperature for a constant value of the curvature}
\end{figure}
\section{Remarks and conclusions}

In this paper we have calculated the one-loop finite temperature
effective potential (or free energy ) for a system of scalar
particles on a manifold with compact hyperbolic spatial part. We
have renormalized  the zero temperature part of the finite
temperature effective potential, and our result  are according with
[13]. In the limit of high temperature and a flat space-time ($R=0$)
we found the well know result of the literature, see for example
[23].

Carefully analyzing the free energy of our model, we realize that
there exist a minimum of the free energy with respect to the
curvature. We call this minimum by $F_{{\small m}}(R_{e},T)$, where
$R_{e}$ is a stationary curvature that minimize the free energy. The
equilibrium state of the system is given by a minimum of the free
energy as a function of the temperature and the stationary
curvature, i.e., that is the statement of the principle of minimum
free energy. The stationary curvature is a function of the
temperature. But, we only can estimate numerically this stationary
curvature for a given value of the temperature. Because the
expression is very complicate. When the temperature increases, the
value of the stationary curvature increases too. Therefore, we can
write the following relation of proportionality for the state of
equilibrium: $R_{e}\propto T$.

This result can be once again emphasized in a very simple way.
Because of the principle of minimum free energy (or maximum of the
entropy in equilibrium situation), the curvature is connected with
the temperature. If we consider the primordial universe with very
high temperature, then the curvature will be very high too. This
model is consistent with a primordial universe with large curvature
and temperature, and another situation where the universe is
evolving into a new situation of low temperature and low curvature.
In other words, for low temperature the universe described by our
model becomes flat.

\section*{Acknowledgements}

This work was supported in part by the CAPES, CNPQ and FAPESP. The
authors would like to thank Andrey Bytsenko, Antônio Edson Gonçalves
and B.M. Pimentel for useful discussions and comments.

\selectlanguage{english}

\end{document}